\documentclass[11pt,twoside]{article}
\usepackage{./asp2014}
\usepackage{abstract}
\usepackage[utf8]{inputenc}
\newcommand{\Msol}{M$_{\odot}$}
\newcommand{\HI}{H\,{\sc {i}}}

\newcommand{\Msold}{M$_{\odot}$\,yr$^{-1}$}
\usepackage{abstract}

\renewcommand{\thefootnote}{\fnsymbol{footnote}}
\aspSuppressVolSlug
\resetcounters

\begin{document}
\title{CO and \HI ~emission from the circumstellar envelopes of some evolved stars 
\thanks{This paper makes use of the following ALMA data: ADS/JAO.ALMA$\#$2011.0.0223.S. ALMA is a partnership of ESO (representing its member states), NSF (USA) and NINS (Japan), together with NRC (Canada) and NSC and ASIAA (Taiwan), in cooperation with the Republic of Chile. The Joint ALMA Observatory is operated by ESO, AUI/NRAO and NAOJ. This research has made use of the SIMBAD and ADS databases and is based on observations carried out with the IRAM Plateau de Bure Interferometer (now called NOEMA) and the IRAM 30 m telescope. IRAM is supported by INSU/CNRS (France), MPG (Germany) and IGN (Spain). The (J)VLA data used for this project were obtained as part of programs AM798 and AM1126.}}
\author{P.N. Diep$^{1*}$, D.T. Hoai$^1$, P.T. Nhung$^1$, P. Tuan-Anh$^1$, T. Le Bertre$^2$, J.M. Winters$^3$, L.D. Matthews$^4$, N.T. Phuong$^1$, N.T. Thao$^1$ and P. Darriulat$^1$\\
$^*$Presenting author: pndiep@vnsc.org.vn
\affil{$^1$Dept of Astrophysics, VNSC/VAST, 18 Hoang Quoc Viet, Hanoi, Vietnam}
\affil{$^2$LERMA, UMR 8112, 61 av. de l'Observatoire, F-75014 Paris, France}
\affil{$^3$IRAM, 300 rue de la Piscine, F-38406 St. Martin d'Hères, France}
\affil{$^4$MIT Haystack Observatory, Off Route 40, Westford, MA 01886, USA}}

\paperauthor{P.N. Diep}{pndiep@vnsc.org.vn}{}{Vietnam National Satellite Center}{Department of Astrophysics}{Hanoi}{Cau Giay/Ha Noi}{10000}{Vietnam}
\paperauthor{D.T. Hoai}{dthoai@vnsc.org.vn}{}{Vietnam National Satellite Center}{Department of Astrophysics}{Hanoi}{Cau Giay/Ha Noi}{10000}{Vietnam}
\paperauthor{P.T. Nhung}{pttnhung@vnsc.org.vn}{}{Vietnam National Satellite Center}{Department of Astrophysics}{Hanoi}{Cau Giay/Ha Noi}{10000}{Vietnam}
\paperauthor{P.T. Anh}{ptanh@vnsc.org.vn}{}{Vietnam National Satellite Center}{Department of Astrophysics}{Hanoi}{Cau Giay/Ha Noi}{10000}{Vietnam}
\paperauthor{T. Le Bertre}{thibaut.lebertre@obspm.fr}{CNRS $\&$ Observatoire de Paris/PSL}{LERMA, UMR 8112}{F-38406 St. Martin d 'Hères}{France}{}{}{}
\paperauthor{J.M. Winters}{winters@iram.fr}{Domaine Universitaire}{IRAM, 300 rue de la Piscine}{F-38406 St. Martin d 'Hères}{France}{}{}{}
\paperauthor{L. Matthews}{lmatthew@haystack.mit.edu}{MIT Haystack Observatory}{Off Route 40}{Westford/MA 01886}{USA}{}{}{}
\paperauthor{N.T. Phuong}{ntphuong02@vnsc.org.vn}{}{Vietnam National Satellite Center}{Department of Astrophysics}{Hanoi}{Cau Giay/Ha Noi}{10000}{Vietnam}
\paperauthor{N.T. Thao}{ntthao02@vnsc.org.vn}{}{Vietnam National Satellite Center}{Department of Astrophysics}{Hanoi}{Cau Giay/Ha Noi}{10000}{Vietnam}
\paperauthor{P. Darriulat}{darriulat@vnsc.org.vn}{}{Vietnam National Satellite Center}{Department of Astrophysics}{Hanoi}{Cau Giay/Ha Noi}{10000}{Vietnam}

\begin{abstract}
Studies of the CO and \HI ~radio emission of some evolved stars are presented using data collected by the IRAM Plateau de Bure interferometer and Pico Veleta telescope, the Nançay Radio Telescope and the JVLA and ALMA arrays. Approximate axial symmetry of the physical and kinematic properties of the circumstellar envelope (CSE) are observed in CO emission, in particular, from RS Cnc, EP Aqr and the Red Rectangle. A common feature is the presence of a bipolar outflow causing an enhanced wind velocity in the polar directions. \HI ~emission extends to larger radial distances than probed by CO emission and displays features related to the interaction between the stellar outflow and interstellar matter. With its unprecedented sensitivity, FAST will open a new window on such studies. Its potential in this domain is briefly illustrated. 
\end{abstract}
\saythanks
\renewcommand{\thefootnote}{\arabic{footnote}}

\section{CO emission}
We present observations of the CO emission of three evolved stars, EP Aqr, RS Cnc and the Red Rectangle (RR). They are representative of three different stages of evolution. EP Aqr, from the absence of Technetium in its CSE, is identified as a young AGB star, RS Cnc is in its thermal pulsing phase and has experienced several dredge-up episodes and the RR is a post-AGB star in the process of becoming a Planetary Nebula. The outflows of all three stars display approximate rotational invariance about an axis but a distinctive feature is the different orientation of this axis with respect to the sky plane: nearly perpendicular for EP Aqr, at about $45^\circ$ for RS Cnc and nearly parallel for the RR. Detailed accounts of the observed features have been recently published \citep{ex_1,ex_2,ex_3}; in the present contribution we shall be satisfied with commenting briefly on the main results.

Reconstruction in space, from the observed spectral maps, of the density, temperature and velocity of the gas envelope is only possible at the price of some simplifying hypotheses such as local thermal equilibrium or rotational symmetry about the star axis. In this context, it is convenient to define an effective density $\rho$ such that $F(y,z)=\int{f(y,z,V_x)dV_x}=\int{\rho(x,y,z)dx}$. Here $y$, $z$ and $x$ point respectively east, north and along the line of sight away from Earth; $f(y,z,V_x)$ is the measured flux density at Doppler velocity $V_x$; $F(y,z)$ is its integral over $V_x$.

All three stars display similar two-component Doppler velocity distributions, symmetric about the origin, providing evidence for winds at the 10 kms$^{-1}$ scale. For RS Cnc and EP Aqr, which have mass loss rates in the range of $10^{-7}$ \Msold, the wind has been insufficient to significantly shape the effective density, which displays approximate circular symmetry in the plane of the sky. On the contrary, in the case of the RR, which has formerly experienced an episode of strong mass loss, the effective density is clearly distorted as seen on the sky maps of $RF(y,z)$, with $R=\sqrt{y^2+z^2}$, displayed in Figure 1. The effect of the wind is best revealed by mapping the mean Doppler velocity $<V_x>$, which shows a clear asymmetry in the RS Cnc and RR cases, associated with expansion for the former and with rotation for the latter (Figure 1).

\begin{figure}[!ht]
\centering
\includegraphics[height=2.7cm]{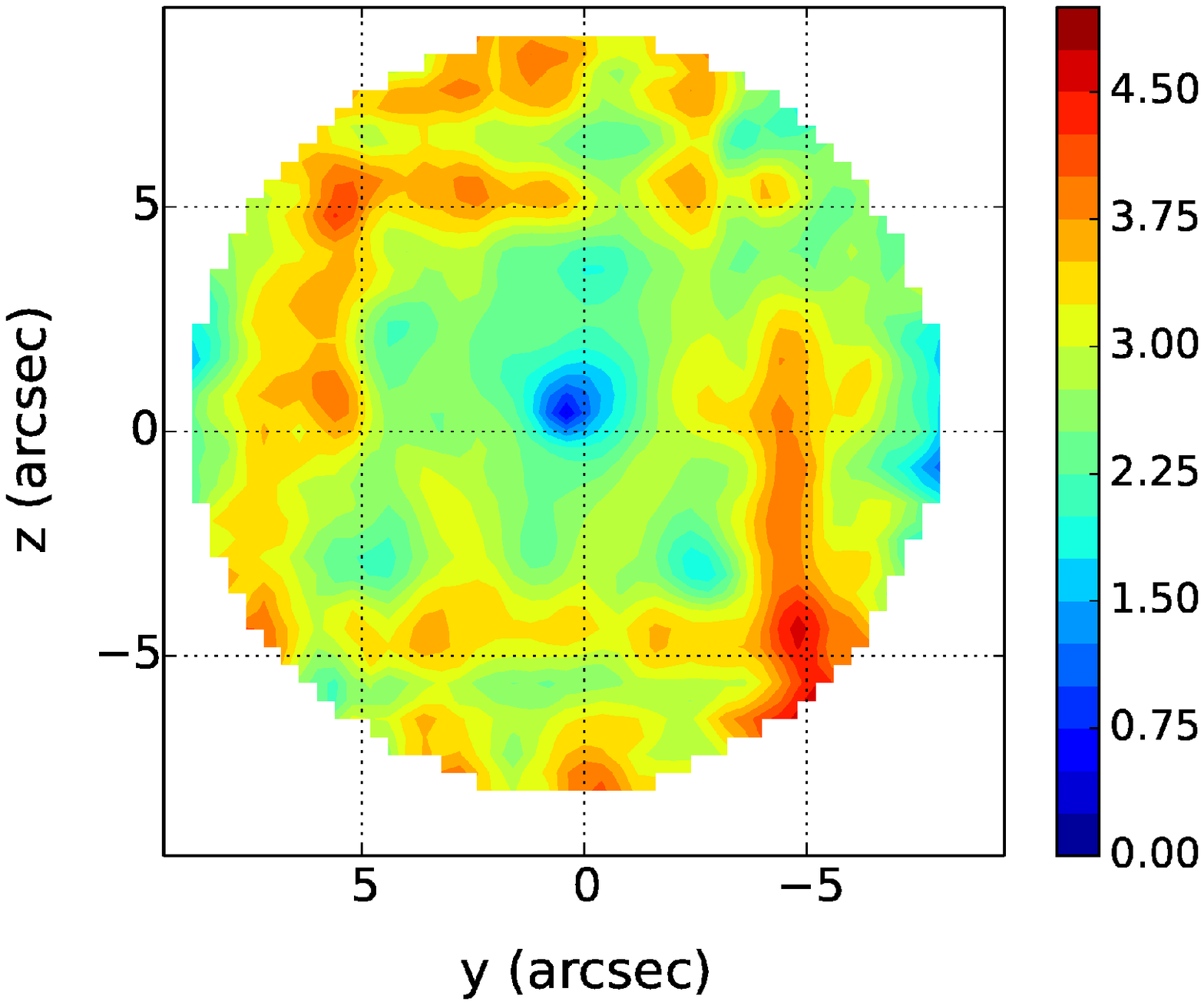}
\includegraphics[height=2.7cm]{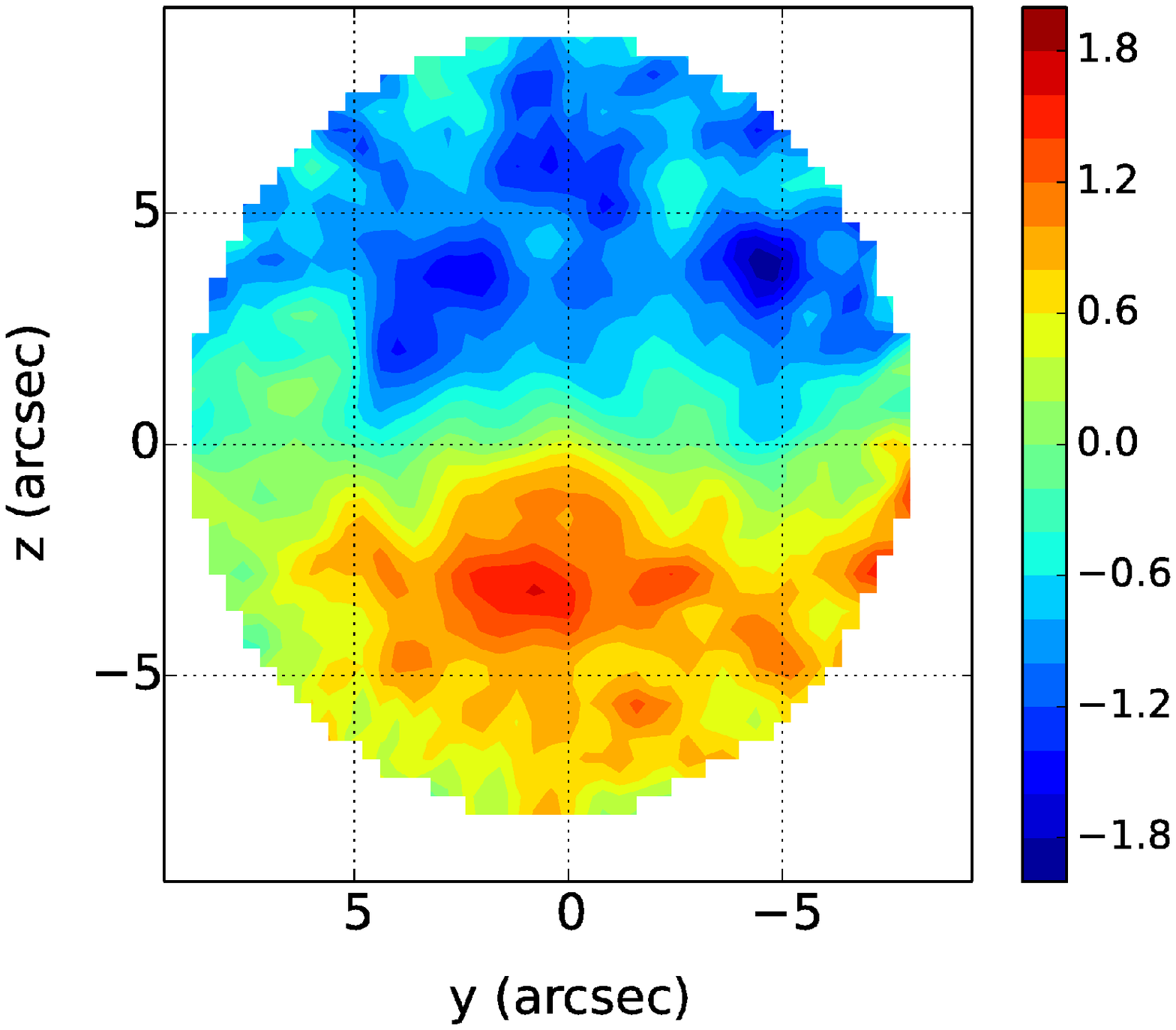}
\includegraphics[height=2.7cm,trim=0.cm -.5cm 0.cm 0.cm,clip]{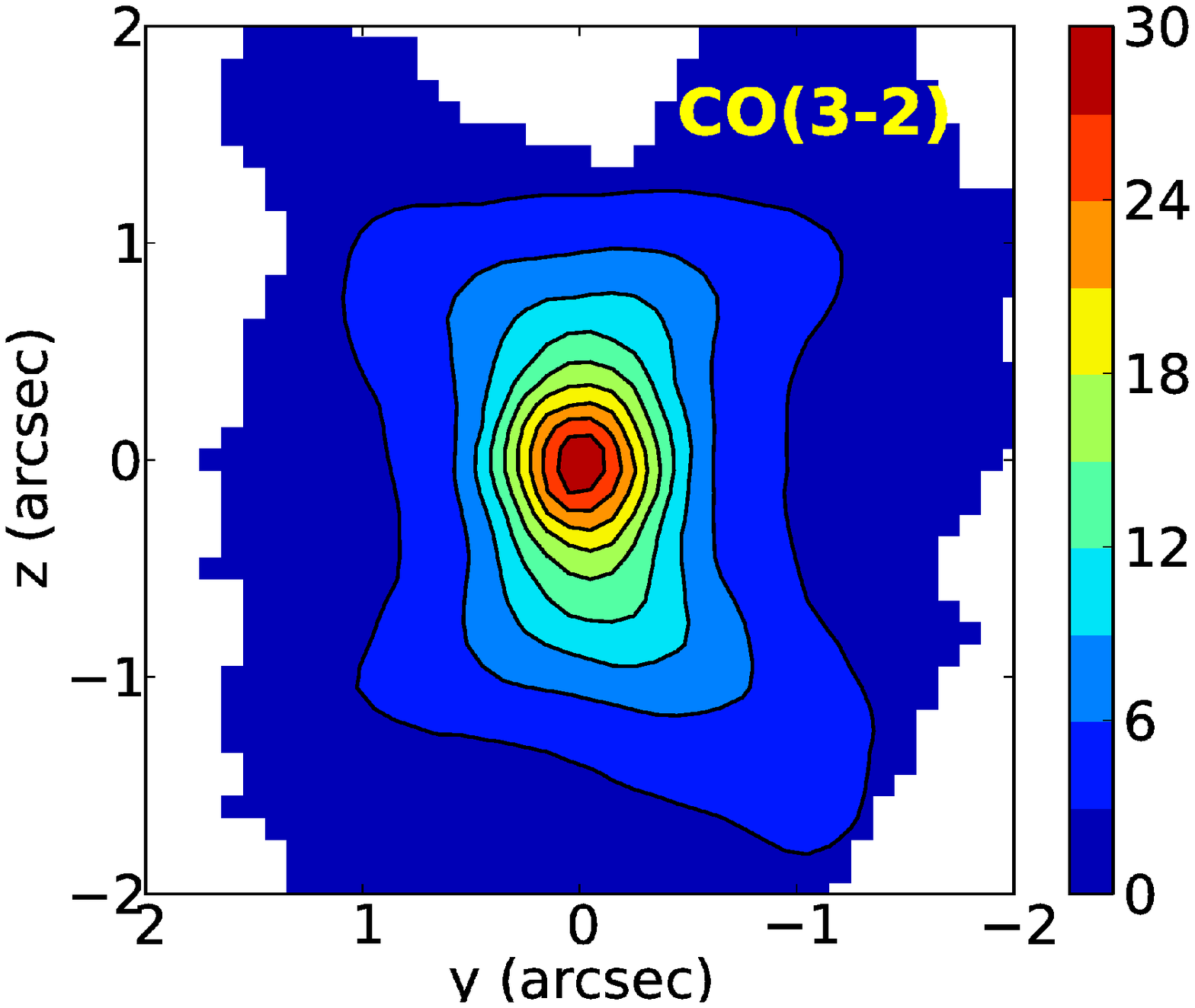}
\includegraphics[height=2.7cm,trim=0.cm -.5cm 0.cm -.2cm,clip]{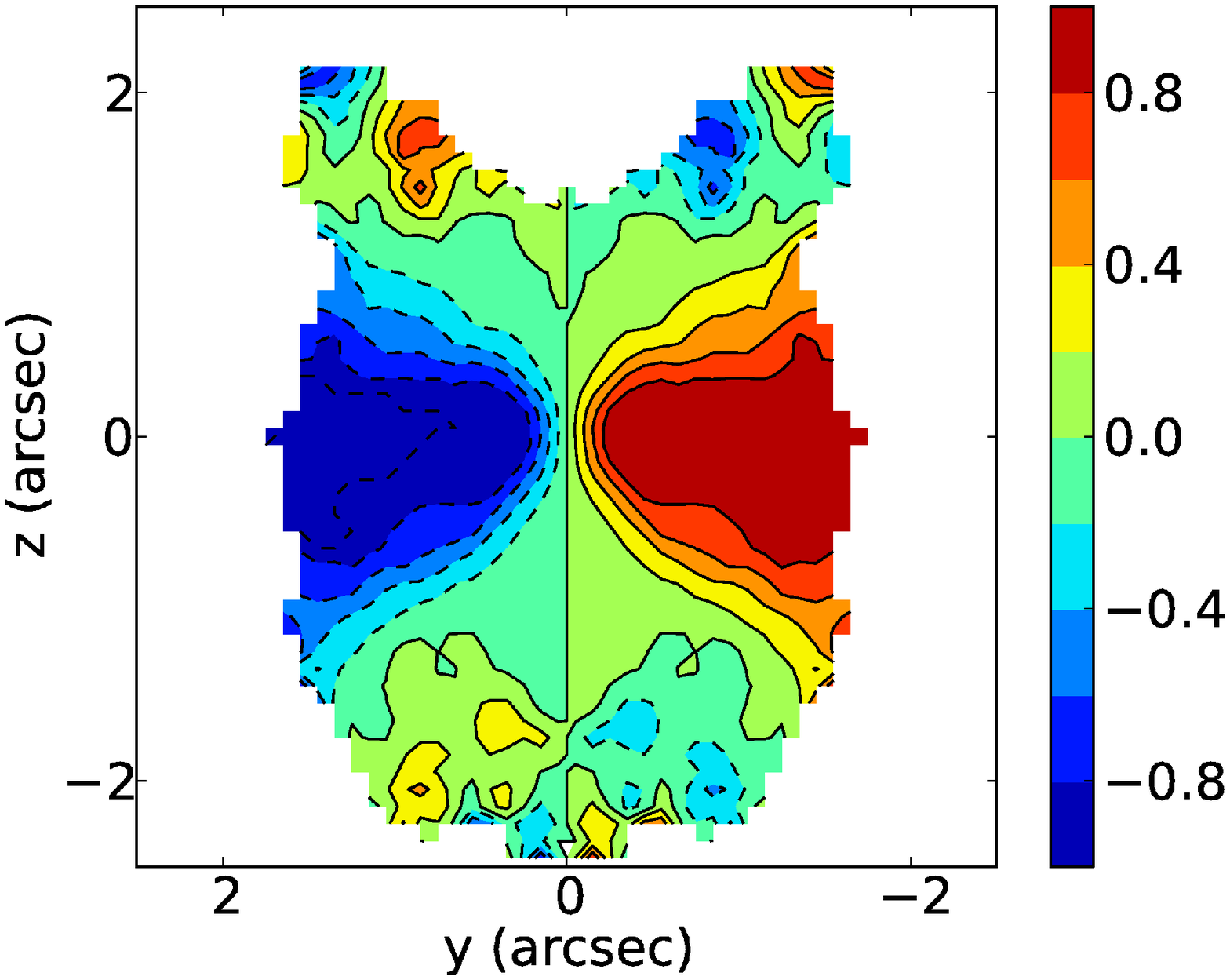}
\caption{Sky maps of CO observations made of RS Cnc (left panels) in CO(1-0) \citep{ex_15} and of the RR (right panels) in CO(6-5) \citep{ex_3}. In each panel, the first shows $RF(y,z)$ and the second $<V_x>$ (or a quantity closely related to it in the RR case).}
\end{figure}

In the case of EP Aqr, no strong asymmetry is revealed, suggesting either isotropic emission \citep{ex_4} or a star axis parallel to the line of sight \citep{ex_1}. Indeed, analyses of its CO emission have been made under both hypotheses, the second hypothesis being however favored over the former.

The peculiar geometries of EP Aqr and of the RR allow for reconstructing in space the gas properties under simplifying hypotheses. In the RR case, the effective density can be reconstructed in space by simply solving the integral equation $F(y,z)=\int{\rho(x,y,z)dx}$ when the effective density is taken invariant by rotation about the star axis. Indeed, this equation does not mix different $z$ values and therefore effectively splits into as many equations as there are values of $z$. For each of these, $\rho(x,y,z)$ depends exclusively on $\xi=\sqrt{x^2+y^2}$ and the integral equation is easily solved.

The situation is very different in the case of EP Aqr. In this case, rotational invariance means circular symmetry on the sky plane and does not help with the solution of the integral equation: the radial and latitudinal dependences of the effective density are mixed. Indeed, there is no way to distinguish between spherical symmetry and axial symmetry about an axis parallel to the line of sight when considering the effective density alone. It is then necessary to make some hypothesis on the form taken by the radial and/or latitudinal dependences of the wind velocity. For example, assuming the wind to be purely radial, the relation $x/R=V_x/\sqrt{V^2-V_x^2}$ provides a direct measurement of $x$ once $V$ is known. It becomes then possible to solve the integral equation.
 
\begin{figure}[!ht]
\centering
\includegraphics[width=.8\textwidth]{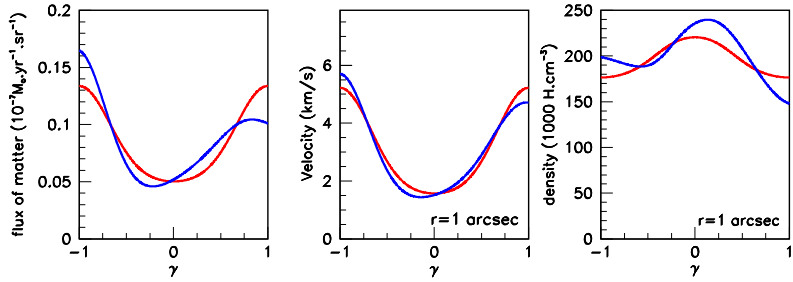}
\caption{RS Cnc best fit results assuming two different versions of the model (centrally symmetric, red, or asymmetric, blue). Dependence on the sine of the stellar latitude of the flux of matter (left), of the wind velocity (middle) and of the gas density at $r$=1'' (right).} 
\end{figure}

In general, however, a model of the effective density and wind velocity needs to be constructed in order to obtain a description of the properties of the gas envelope in space. Such a model includes radiative transfer and assumes the wind to be purely radial, free of turbulence and in local thermal equilibrium. Moreover, it is supposed to have been in such a regime for long enough a time, such that the radial extension of the gas volume is governed exclusively by the UV dissociation of the CO molecules by interstellar radiation \citep{ex_5} and does not keep any trace of the stellar mass-loss history. The temperature is parameterized as functions of $r=\sqrt{x^2+y^2+z^2}$ and latitude and requires the observation of at least two molecular lines of the same species in order to be constrained by the data. Both the effective density and the radial wind velocity (and therefore the flux of matter, allowance being made for the presence of velocity gradients) are assumed to vary smoothly from equator to poles. Flux densities are calculated in each pixel by integration along the line of sight, the temperature dependent contributions of emission and absorption being respectively added and subtracted at each step. For both RS Cnc and EP Aqr, as well as for several other AGB stars not mentioned here, a satisfactory description of the observed spectral maps has been obtained by adjusting the model parameters to best fit CO(1-0) and CO(2-1) emissions together. For stars having a mass loss rate not exceeding a few $10^{-7}$ \Msold, the effective density is found not to strongly deviate from spherical and the wind velocity to vary from values at the kms$^{-1}$ scale at the equator to the 10 kms$^{-1}$ scale at the poles (Figure 2). There is of course some arbitrariness in the results obtained, their validity resting on assumptions such as the wind being radial and the properties of the gas envelope being invariant about the star axis. However, while important deviations from such an idealized scenario are revealed, they are shown not to strongly affect the general picture.  

Comparing CO(1-0) and CO(2-1) emission provides an evaluation of the temperature as a function of stellar latitude and distance from the central star. Both RS Cnc and EP Aqr display an equatorial temperature enhancement (Figure 3). In the RR case, the high spatial resolution allows for a more detailed description of the morphology and kinematics, with evidence for a clear separation between a rotating equatorial gas volume and conical polar outflows. Comparing CO(3-2) and CO(6-5) emissions reveals a temperature enhancement in the polar outflows, and particularly at the interface with the equatorial rotating volume (Figure 3). 

\begin{figure}[!ht]
\centering
\includegraphics[height=4.cm,trim=0.cm .12cm 0.cm 0.cm,clip]{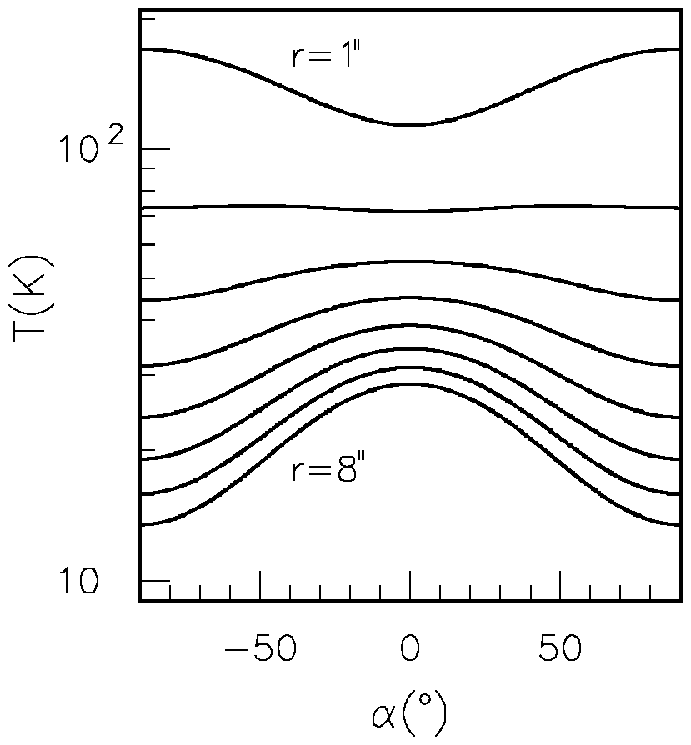}
\includegraphics[height=4.1cm]{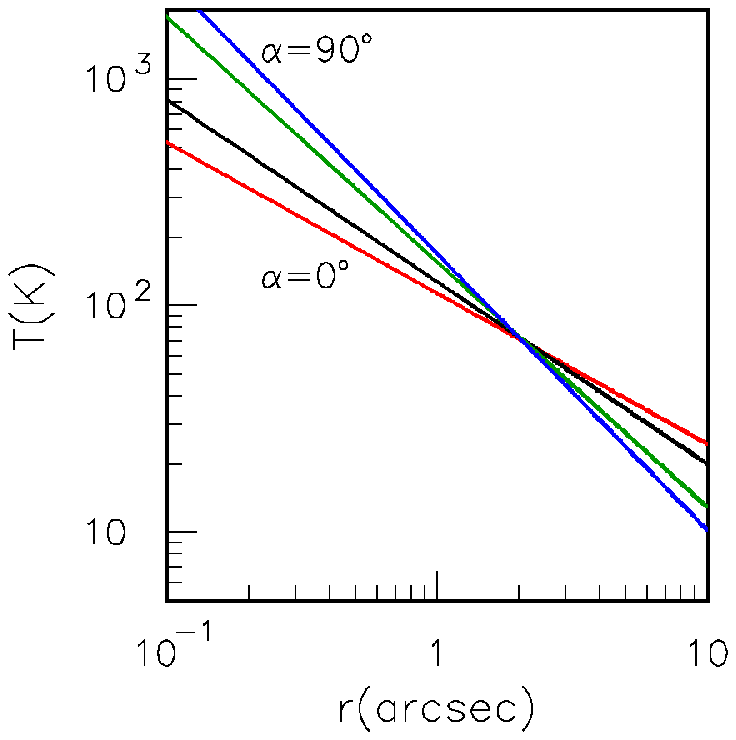}
\includegraphics[height=4.1cm,trim=0.cm -1.8cm 0.cm 0.cm,clip]{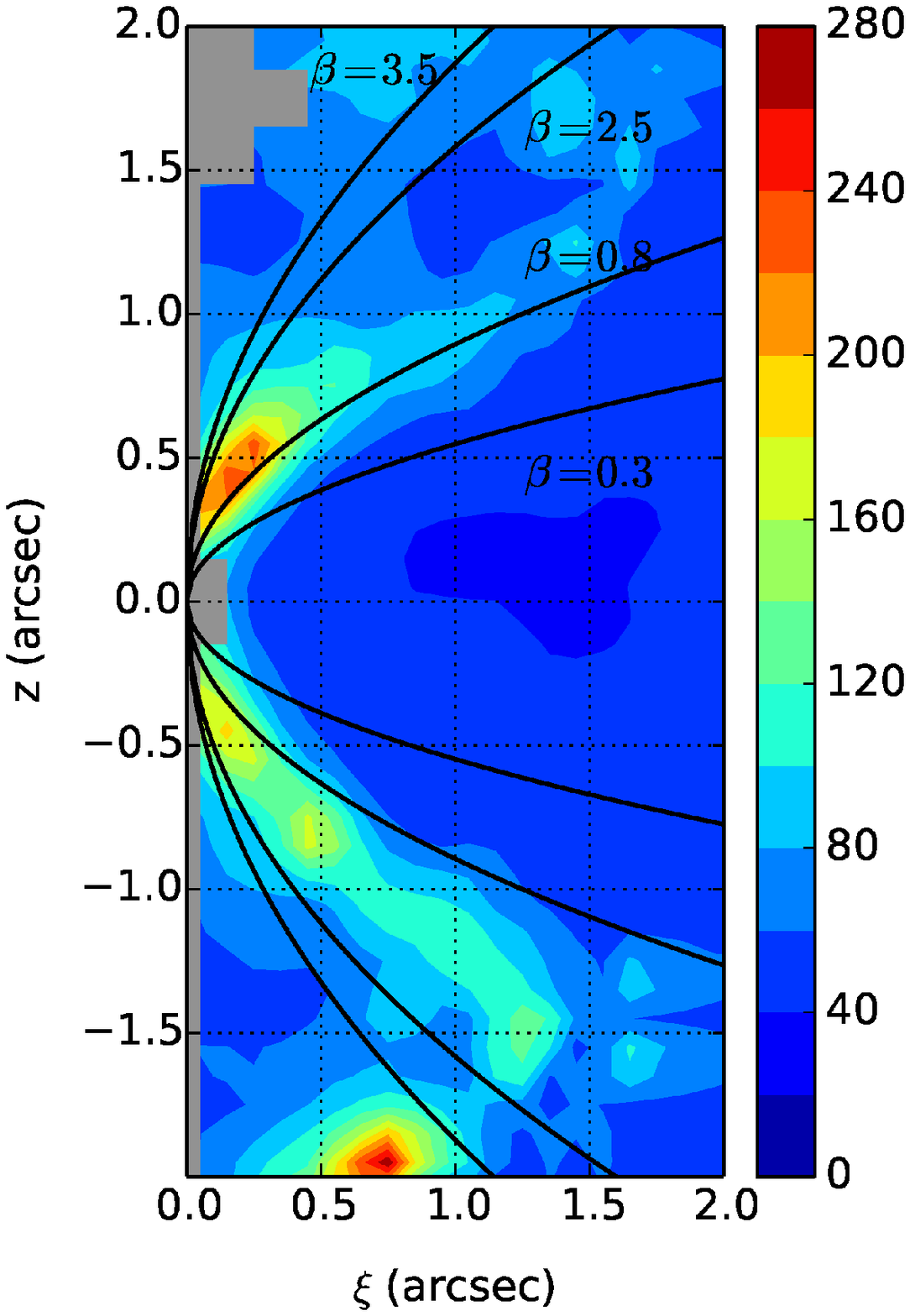}
\caption{Temperature dependence on stellar latitude $\alpha$ and on $r$. Left and middle panels are for EP Aqr as a function of $\alpha$ at $r$ = 1'' to 8'' (from top to bottom) in steps of 1'' and as a function of $r$ at latitudes $\alpha=0^{\circ}$ (red), $30^{\circ}$ (black), $60^{\circ}$(green) and $90^{\circ}$ (blue). The right panel maps the temperature in the meridian plane of the RR.}
\end{figure}

\section{\HI ~emission}

The use of carbon monoxide as a tracer is limited to the inner parts of the circumstellar shells because, at a distance of some $10^{17}$ cm, it is dissociated by the interstellar radiation field \citep{ex_5}. At larger distances, it is necessary to use other tracers, such as dust or atomic species. The 21 cm \HI ~line is very good in this role \citep{ex_6}. Both EP Aqr and RS Cnc have been the objects of pioneering high spectral resolution \HI ~detection using the upgraded Nançay Radio Telescope \citep{ex_7,ex_8} and followed by high spatial resolution observations using the Very Large Array \citep{ex_9}.

\begin{figure}[!ht]
\centering
\includegraphics[height=4.3cm]{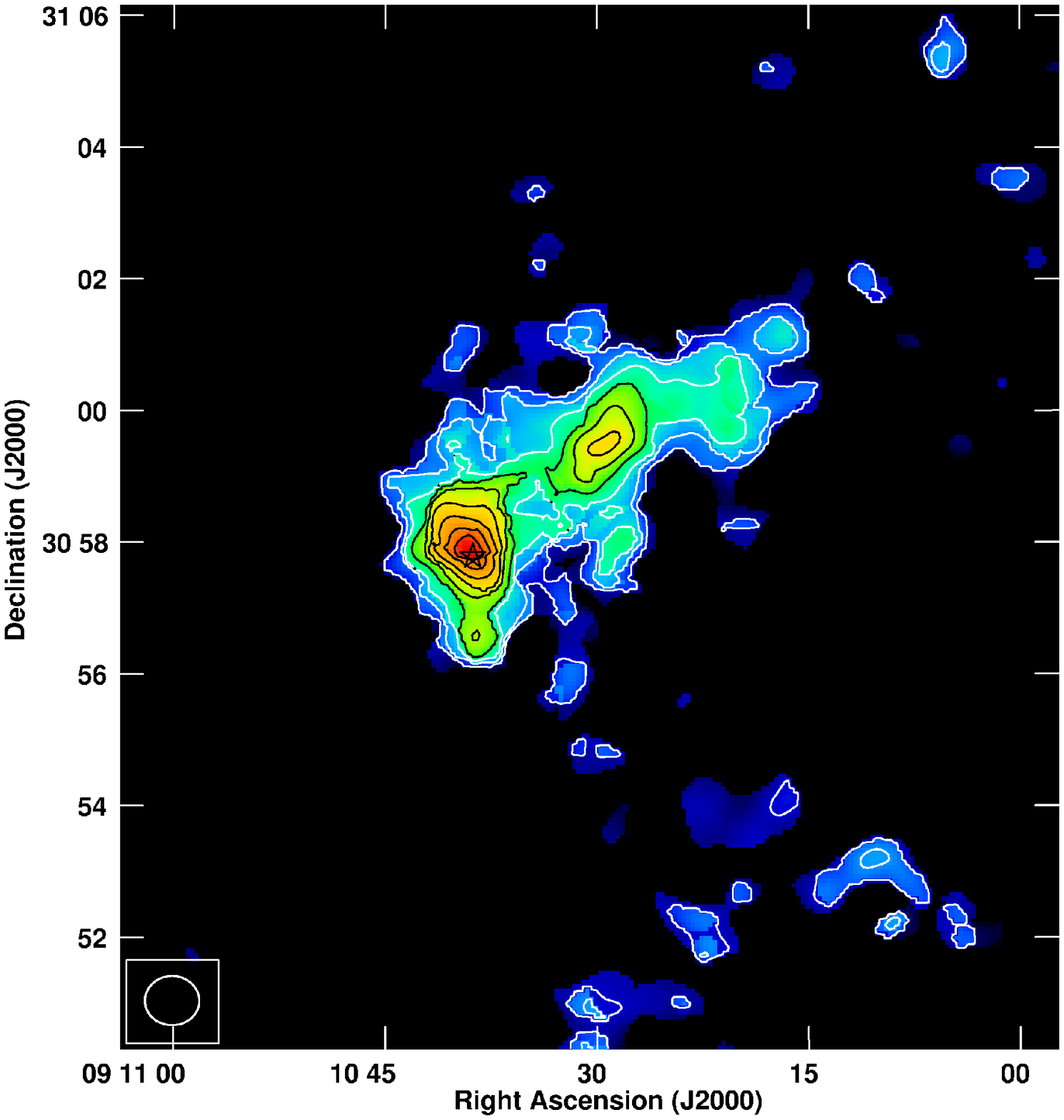}
\includegraphics[height=4.25cm]{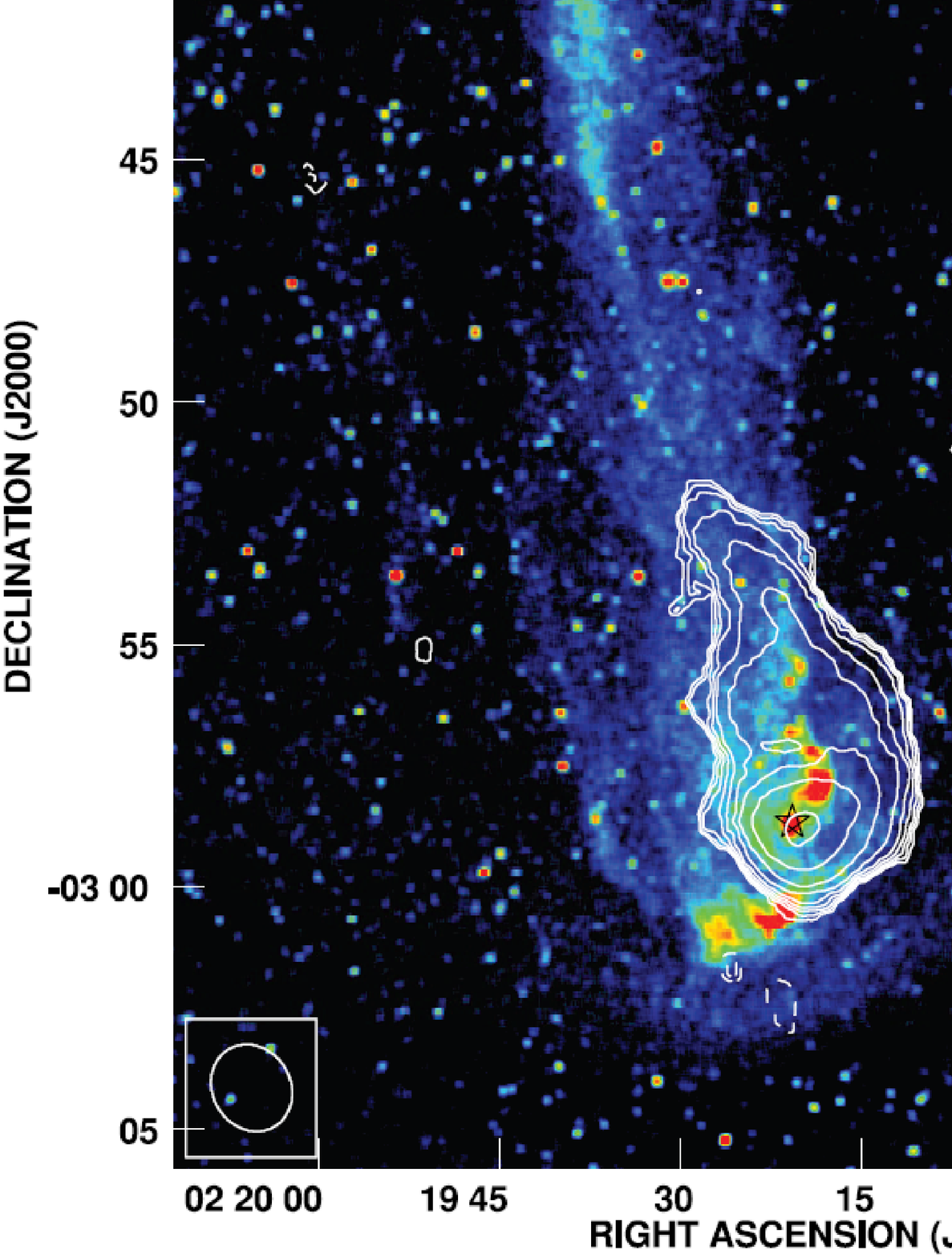}
\caption{\HI ~total intensity maps of RS Cnc (left) and Mira Ceti (right, shown as contours overlaid on a GALEX FUV image).}
\end{figure}

The \HI ~map of RS Cnc (Figure 4, left) \citep{ex_15} shows a 6'-long tail, in a direction opposite to the space motion of the central star, and a compact structure of size $\sim75''\times40''$ centered near the star position, the ``head''; it is elongated in the direction of the bipolar outflow observed in CO at shorter distances, up to $\sim 10''$, suggesting that \HI ~is tracing an extension of the molecular outflow, interacting with the local interstellar medium, beyond the CO dissociation radius. However, the velocity range of the detected \HI ~emission is significantly smaller than that of CO emission, suggesting either that the atomic hydrogen has been slowed down in its interaction with the interstellar medium or that it was emitted at an earlier epoch when the bipolar outflow was not yet important. The velocity-integrated \HI ~flux density is measured as $1.14\pm0.03$ Jykms$^{-1}$ implying a mass of $\sim5.5\times 10^{-3}$ \Msol.

At variance with Mira Ceti (Figure 4, right) \citep{ex_11} and X Her \citep{ex_12}, two AGB stars displaying \HI ~tails analogous to RS Cnc, no velocity gradient is detected along the RS Cnc tail but evidence for a complex structure of the velocity field is revealed by the presence of important inhomogeneities.

While a clear \HI ~signal was detected around EP Aqr by the Nançay Radio Telescope, subsequent observation using the VLA revealed the presence of multiple clumps redward of the star velocity (while CO emission is well centered). Added together, their peak flux density reaches $\sim 24$ mJy. The quality of the data does not allow for an unambiguous identification of the observed \HI ~emission with EP Aqr mass loss, confusion with galactic sources along the line of sight being possible.

Finally, to our knowledge, no \HI ~emission has ever been observed around the Red Rectangle or around post-AGB stars having mass loss rates much higher than AGB stars such as EP Aqr and RS Cnc. Several sources with high mass loss rates, such as IK Tau or AFGL 3099, which are observed at high galactic latitudes where the background and foreground interstellar \HI ~emission is not a problem, remain undetected \citep{ex_6,ex_13}. A simple model simulation, in the spirit of what was described for CO emission, shows that neither background absorption nor optical thickness can explain it \citep{ex_10}. Recently, \cite{ex_14} have reported the detection of atomic hydrogen in the circumstellar environment of IRC+10216, a late AGB star having a high mass loss rate of the order of $2\times10^{-5}$ \Msold. They find that atomic hydrogen accounts for less than $1\%$ of the expected circumstellar mass, supporting a composition dominated by molecular hydrogen. This might then be true for other high mass loss rate stars, explaining the absence of \HI ~detection in their environment. 

\section{Future expectations}
In both the CO and \HI ~cases, recent progress has been spectacular and calls for data of even better quality in order to answer the many question marks that persist. In the former case, ALMA is opening a new window on the morphology and kinematics of the CSE at small distances from the star. In the latter case, the need for higher surface brightness sensitivities, while maintaining good spectral and spatial resolutions, is clear. With FAST coming up, important progress will be at hand along this line. We have developed a radiative transfer model that will allow us to interpret the data that FAST will obtain on circumstellar shells \citep{ex_10}.

\HI ~observations are unique in measuring gas velocities at large distances from the star. Together with dust observations in the infrared, they allow for mapping the gas volume on the sky. Despite the abundant motivations for measuring the hydrogen component in the CSEs of AGB stars, such observations are challenging in practice. The \HI ~signal is weak and often competes with strong Galactic background and foreground emission along the line of sight. Until now around some two dozens of AGB stars have been detected in \HI: the emission is weak with narrow line profiles (a few kms$^{-1}$). The sizes of the sources at less than 500 pc are a few arcminutes to one degree. \HI ~reveals external parts of circumstellar shells where stellar winds are interacting with the local ISM and where the material is mostly atomic. \HI ~is a unique morpho-kinematic tracer of these regions but, as we have seen, many sources with high mass loss rate remain undetected. Current issues that can be addressed in \HI ~are to probe the morphology of the outer circumstellar shells and their physical conditions; to constrain the H${_2}$/\HI ~conversion mechanism and the mass loss history; to observe the injection of circumstellar matter into the ISM and; to test and constrain hydrodynamical models describing the interaction of circumstellar shells with the ISM. 

With its high sensitivity, and high mapping efficiency, FAST will be very well suited for performing surveys of the Galactic plane, that could be used to identify counterparts in \HI ~of red giant stars with high mass loss rates (e.g. OH/IR stars), which up to now have been elusive. In addition, \HI ~sources that are close enough to map their spatio-kinematic structure with FAST include IRC+10216, RS Cnc, and Mira (all of which have distances <500 pc).

Assuming a gain of 16 K/Jy and a system temperature of 30 K, a $3\sigma$-sensitivity of 10 mJy in a channel of 1 kms$^{-1}$ should be reached in $\sim5$ minutes. With 19 beams of 2.4 arcminutes, a field of $0.5^{\circ}\times3^{\circ}$ could be covered in $\sim5$ hours. As a pilot program, a high sensitivity observation of Mira Ceti's tail, which cannot be observed by the Arecibo telescope, would be ideal for evaluating the performances of FAST, while providing a unique insight on the kinematics of the gas at large distances from the source.

\acknowledgements We are indebted and very grateful to the ALMA partnership, who are making their data available to the public after a one year period of exclusive property, an initiative that means invaluable support and encouragement for Vietnamese astrophysics. We particularly acknowledge friendly and patient support from the staff of the ALMA Helpdesk. Financial support from the World Laboratory, from the French CNRS (Vietnam/IN2P3 LIA and ASA), from FVPPL, from PCMI, from VNSC, from NAFOSTED and from the Rencontres du Vietnam is gratefully acknowledged. LDM is supported by grant 6928655 from the National Science Foundation.
            



\end{document}